\documentclass{elsart41}
\usepackage{graphics}
\usepackage{graphicx}
\usepackage{amssymb}
\begin{document}

\begin{frontmatter}



\title{The effects of the symmetric and antisymmetric anisotropies
on the dynamics of the spin-$\frac{1}{2}$ $XY$ chain}
%---- Don't remove this comment line! ----

\author[ua]{O. Derzhko},
\author[ua]{T. Verkholyak\corauthref{Verkholyak}},
\ead{werch@icmp.lviv.ua}
\author[ua]{T. Krokhmalskii},
\author[de]{H. B\"{u}ttner}

\address[ua]{Institute for Condensed Matter Physics NASU, 
1 Svientsitskii Str., L'viv-11, 79011, Ukraine}  
\address[de]{Theoretische Physik I, Universit\"{a}t Bayreuth,
Bayreuth, D-95440, Germany}

\corauth[Verkholyak]{Corresponding author. 
Tel: +38 0322 761978,
fax: +38 0322 761158}

\begin{abstract}

The dynamic properties of the spin-$\frac{1}{2}$ anisotropic $XY$ chain 
with the Dzyaloshinskii-Moriya (DM) interaction 
in a transverse field 
are investigated.
Using the Jordan-Wigner transformation,
the dynamic structure factors of the model are evaluated rigorously 
(partially analytically and partially numerically). 
The effects of the DM interaction 
on the frequency shapes of the dynamic structure factors 
are discussed.

\end{abstract}

\begin{keyword}
quantum spin chains 
\sep 
Dzyaloshinskii-Moriya interaction 
\sep 
dynamic structure factors
% PACS codes here, in the form: 
\PACS 75.10.Jm
\end{keyword}
\end{frontmatter}

% main text

%%%%%%%%%%%%%%%%%%%%%%%%%
One-dimensional magnetic systems
exhibit a variety of interesting phenomena 
which are the subject of intensified theoretical and experimental studies nowadays.
In this paper,
we study the dynamic properties 
of the spin-$\frac{1}{2}$ anisotropic $XY$ chain 
with the Dzyaloshinskii-Moriya (DM) interaction directed along $z$-axis in spin space 
in a transverse 
(i.e. parallel to $z$-axis in spin space) 
magnetic field.
The one-dimensional spin-$\frac{1}{2}$ $XY$ model 
is related to some quasi-one-dimensional compounds 
(e.g. Cs$_2$CoCl$_4$ \cite{bib1}).
On the other hand,
the DM interaction is often present 
in the models of many low-dimensional magnetic materials.
Despite being small,
this interaction is known to give rise 
to many spectacular features of such compounds.
However, 
we do not intend to fit the spin model in question 
to some specific real systems.
The merit of the considered model is exact solvability, 
i.e. the possibility to obtain for this model reliable conclusions 
avoiding different uncontrolled approximations.

To be specific,
we consider $N\to\infty$ spins one-half 
governed by the Hamiltonian
\begin{eqnarray}
H=\sum_{n}
\left(
J^xs_n^xs_{n+1}^x+J^ys_n^ys_{n+1}^y
\right.
\nonumber\\
\left.
+D\left(s_n^xs_{n+1}^y-s_n^ys_{n+1}^x\right)
\right)
+\sum_{n}\Omega s_n^z.
\label{1}
\end{eqnarray}
This model was introduced in \cite{bib2,bib3}.
Some of its dynamic properties were examined in \cite{bib4,bib5,bib6,bib7}.
After applying the Jordan-Wigner transformation 
the spin system  (\ref{1}) 
can be mapped onto a system of noninteracting spinless fermions.
In our calculations we impose both periodic and open boundary conditions for the spin model (\ref{1})
bearing in mind 
that boundary conditions are irrelevant in the thermodynamic limit 
for bulk characteristics. 
The spinless fermions 
which represent the model (\ref{1}) on a ring 
are governed by the Hamiltonian
\begin{eqnarray}
H=\sum_{\kappa}\Lambda_{\kappa}
\left(\beta_{\kappa}^{\dagger}\beta_{\kappa}-\frac{1}{2}\right),
\;
\Lambda_{\kappa}=D\sin\kappa+\lambda_{\kappa}
\label{2}
\end{eqnarray}
with
$\lambda_{\kappa}=\sqrt{\left(\Omega+J\cos\kappa\right)^2+\gamma^2\sin^2\kappa}$.
Here 
$-\pi\le\kappa<\pi$ denotes the quasimomentum
which parameterizes the fermions,
$J=\frac{1}{2}\left( J^x+J^y \right)$,
$\gamma=\frac{1}{2}\left( J^x-J^y \right)$.
From Eq. (\ref{2}) one immediately concludes
that the energy spectrum becomes gapless 
when 
{\it {i)}} 
$\gamma^2\le D^2$ and $\Omega^2\le J^2+D^2-\gamma^2$
or
{\it {ii)}} 
$\gamma^2>D^2$ and $\Omega^2=J^2$.

%%%%%%%%%%%%%%%%%%%%%%%%%
In our calculations of the $zz$ dynamic structure factor
we follow the standard route 
(see, e.g., \cite{bib8})
obtaining as a result
\begin{eqnarray}
\label{3}
S_{zz}(\kappa,\omega)
=
\sum_{j=1}^3
\int_{-\pi}^{\pi}{\rm{d}}\kappa_1
B^{(j)}
C^{(j)}
\delta\left(\omega-E^{(j)}\right),
\end{eqnarray}
where
$B^{(1)}
=B^{(3)}
=\frac{1}{4}\left(1-f\right)$,
$B^{(2)}
=\frac{1}{2}\left(1+f\right)$,
$f$ is a certain function which depends on 
$\kappa_1$, $\kappa$, $J$, $\gamma$, $\Omega$
(but not $D$)
(see Eq. (4.3c) of \cite{bib8}),
$C^{(1)}
=
\overline{n_{\kappa_1-\frac{\kappa}{2}}}\,
\overline{n_{-\kappa_1-\frac{\kappa}{2}}}$,
$C^{(2)}
=
\overline{n_{\kappa_1-\frac{\kappa}{2}}}\,
n_{\kappa_1+\frac{\kappa}{2}}$,
$C^{(3)}
=n_{\kappa_1+\frac{\kappa}{2}}n_{-\kappa_1+\frac{\kappa}{2}}$,
$n_\kappa=\left(\exp\left(\beta\Lambda_{\kappa}\right)+1\right)^{-1}$ is the Fermi function,
$\overline{n_{\kappa}}=1-n_{\kappa}$,
$E^{(1)}
=\Lambda_{\kappa_1-\frac{\kappa}{2}}+\Lambda_{-\kappa_1-\frac{\kappa}{2}}$,
$E^{(2)}
=\Lambda_{\kappa_1-\frac{\kappa}{2}}-\Lambda_{\kappa_1+\frac{\kappa}{2}}$,
$E^{(3)}
=-\Lambda_{\kappa_1+\frac{\kappa}{2}}-\Lambda_{-\kappa_1+\frac{\kappa}{2}}$.
In the limit of isotropic interaction $\gamma=0$ 
Eq. (\ref{3}) yields the result reported earlier \cite{bib7}.
In the limit $D=0$ and $T=0$ 
Eq. (\ref{3}) coincides with the expression obtained in \cite{bib8}.
In this case 
(and more generally for $D^2<\gamma^2$)
only the two-fermion excitation continuum
which corresponds to $j=1$
contributes in Eq. (\ref{3}).
In the case $D^2>\gamma^2$ and $T=0$
(or $T>0$)
all three two-fermion excitation continua 
(which correspond to $j=1,2,3$)
come into play in Eq. (\ref{3}).
The properties of the two-fermion excitations
which govern $zz$ dynamics 
of the model (\ref{1}) with $D=0$ 
were elaborated in \cite{bib9,bib8}.
The DM interaction essentially affects 
the two-fermion excitation continua;
we will postpone a complete analysis of this issue to a later paper.

%%%%%%%%%%%%%%%%%%%%%%%%%
In our calculation of the $xx$ and $yy$ dynamic structure factors 
we follow the route described in \cite{bib10}.
Considering a chain of $N=400$ sites 
we first compute the time dependent correlation functions 
$\langle s_j^{\alpha}(t)s_{j+n}^{\alpha}\rangle$
and then perform the Fourier transformations 
with respect to the time and space variables.
To be sure that our results pertain to the thermodynamic limit 
we assess the finite-size effects 
performing many simulations similar to the ones described in \cite{bib10}.

%%%%%%%%%%%%%%%%%%%%%%%%%
We demonstrate the effects of the (weak) DM interaction 
on the dynamics of (weakly) anisotropic $XY$ chain 
at low temperatures 
plotting in Fig. 1
the dependences $S_{\alpha\alpha}(\kappa,\omega)$ vs. $\omega$ 
at $\kappa=0$ and $\kappa=\pi$. 
\begin{figure}[!ht]
\begin{center}
\includegraphics[width=0.35\textwidth]{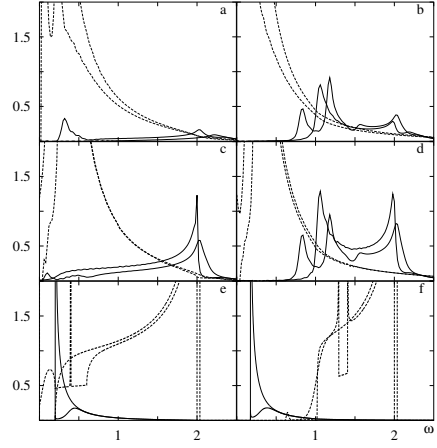}
\end{center}
\caption{The frequency profiles of $xx$ (a,b), $yy$ (c,d) and $zz$ (e,f) dynamic structure factors
at $\kappa=0$ (solid lines)
and $\kappa=\pi$ (dashed lines)
for the spin chain (\ref{1}) 
with $J=1$, 
$\gamma=0.1$, 
$D=0$ (thin lines),
$D=0.2$ (bold lines),
$\Omega=0$ (a,c,e),
$\Omega=0.5$ (b,d,f)
at low temperature $\beta=50$.}
\label{fig1}
\end{figure}
$S_{zz}(\kappa,\omega)$ may have nonzero values 
only within a restricted frequency range 
in correspondence with the two-fermion excitation continua boundaries;
moreover,
it may diverge 
owing to the divergent density of two-fermion states
(panels e, f).
$S_{xx}(\kappa,\omega)$
and
$S_{yy}(\kappa,\omega)$ are not restricted to the two-fermion excitations 
and involve excitations of many fermions.
However,
similarly to the cases 
$\gamma=D=0$ \cite{bib10}
and 
$\gamma=0$, $D\ne 0$ \cite{bib7}
their values are rather small outside the two-fermion excitation continua 
and are concentrated mainly along several 
washed-out excitation branches 
(sharp peaks in panels a - d).
Our results may be of interest for interpreting of the data 
measured by neutron scattering or ESR techniques.
Thus,
the ESR absorption spectrum 
in Faraday configuration
is related to   
$S_{xx}(\kappa,\omega)$,
$S_{yy}(\kappa,\omega)$
\cite{bib11}.
The presented in Fig. 1 frequency shapes clearly demonstrate 
that the DM interaction can change dramatically the dynamic quantities 
observed experimentally 
on materials modeled by the spin-$\frac{1}{2}$ $XY$ chain. 

%%%%%%%%%%%%%%%%%%%%%%%%%
T. V. acknowledges the kind hospitality 
of the University of Bayreuth in the spring and autumn of 2004.
The paper was presented partially  
at the International Workshop on
Collective quantum states in low-dimensional transition metal oxides 
(Dresden, February 22-25, 2005).
O. D. thanks the MPIPKS, Dresden 
for the hospitality.

%%%%%%%%%%%%%%%%%%%%%%%%%

\end{document}